\documentclass{article}
\usepackage{tikz}
\usetikzlibrary{matrix,positioning,quotes}
\usepackage{graphicx} 
\usepackage{amsmath}
\usepackage{mathtools}
\usepackage{float}
\usepackage{subcaption}
\usepackage{url}
\usepackage{pgfplots}
\usepackage{xcolor}

\begin{document}

\title{Population dynamics in the global coral–symbiont network under temperature variations}

\author{\large Maria Gabriella Cavalcante Basílio$^1$\& Daniel Ratton Figueiredo$^1$ \\
	\small $^1$Systems Engineering and Computer Science (PESC),\\
	\small Federal University of Rio de Janeiro (UFRJ),\\
	\small Rio de Janeiro, RJ, Brazil \\
	\large \texttt{\{basiliog,daniel\}@cos.ufrj.br} }

\date{}

\maketitle 

\begin{abstract}
Coral reefs are crucial to marine biodiversity and rely on a delicate symbiotic relationship between corals and zooxanthellae algae. Water temperature variations, however, disrupt this association, leading to coral bleaching events that severely affect marine ecosystems. This study presents a mathematical model for the population dynamics of coral and symbiont species considering the coral–symbiont network and recurrent warming events. The model incorporates thermal tolerances of species and coupled growth dynamics (between corals and symbionts) to investigate how network structure and thermal tolerance influence the species' growth. Using real data from different ocean regions, results reveal that network connectivity plays a significant role in population growth after successive warming events, with generalist species demonstrating greater growth across all regions analyzed. The comparatively higher correlation between node degree and final population also emphasizes the impact of ecological network structure on species growth, offering valuable insights into coral reef population dynamics under climate change. This research highlights the need to consider network structure beyond species' thermal tolerances when evaluating the ecological responses of corals to environmental changes.

\end{abstract}
\section{Introduction}  

\hspace{5mm}The symbiotic relationship between species of coral and microalgae, known as zooxanthellae, is fundamental to the survival of coral reefs, as algae and corals exchange nutrients with each other. However, when these organisms are exposed to warming events, the bond between the host coral and endosymbiotic algae breaks down. The breaking of this bond is called coral bleaching \cite{donner,williams} because the color of corals often depends on the type of algae associated with them. Therefore, when the association is broken, the coral loses its color and becomes white, a phenomenon known as bleaching. 

The bleaching event causes several ecological impacts such as a decrease in its growth rate. Furthermore, bleaching events are recurrent and have become more frequent in the last decade \cite{hughes}. However, warming events do not equally affect all organisms, as they have different thermal tolerances which represents their capacity to grow under water temperature variation \cite{williams,will_tese}. Finally, the capacity to grow also depends on the symbiotic relationships of the organism. Intuitively a coral with multiple and diverse algae has a higher tolerance to warming events and thus more robust growth.  

This work presents a mathematical model for the growth of corals and algae considering both their symbiotic relationship and thermal resistance under recurrent warming events. In particular, a bipartite network encodes the symbiotic relationships and a differential equation for each organism (coral and algae) captures its population growth. This equation depends on the network structure, thermal tolerances, water temperature and the population of the symbionts. 

Using real data collected from different ocean regions, the proposed growth model is applied to a representative model for recurrent warming events. Starting from identical initial populations, the model shows that different organisms have very different growth patterns over time. The relationship between population and network structure and thermal tolerance is investigated. Results indicate that, for different ocean regions, correlation between network structure and population is stronger than thermal tolerance and population. This highlights the importance of the symbiotic network on understanding bleaching events. 

\section{Related work}

\hspace{5mm}Understanding the consequences of rising ocean temperatures in the development of coral reefs through network analysis has been broadly explored in the recent literature. A notable work studies the global network between coral species and \textit{Symbiodiniaceae} and its resistance to temperature stress as well as its robustness to temperature perturbations \cite{williams}. Another recent work proposes and evaluates an eco-evolutionary model that shows that shortcuts in the dispersal network (e.g., corals that disperse larvae throughout the ocean to coral reefs) across environmental gradients (i.e,  changes in non-living factors through space or time) hinder the persistence of population growth across regions \cite{mcmanus,mcmanus2}. Theses works have been quite successful in identifying how the network structure affects the sensitivity of corals to changes in water temperature, either in symbiotic associations networks or in dispersal networks.

For instance, in the context of the global coral-symbiont network \cite{williams}, null networks were created by altering physiological parameters of organisms or the network structures. A bleaching model was developed with weighted links representing temperature thresholds for host–symbiont pairs. Resistance to temperature stress and ecological robustness were assessed by analyzing how different networks responded to increasing temperatures (e.g., link removal) and species (e.g., node) removal. Results indicated that robustness to bleaching and other perturbations varied across spatial scales and differed from null networks. The global coral–symbiont network was more sensitive to environmental attacks, such as rising temperatures, with symbionts providing more stability than hosts. Network structure and thermal tolerances are not represented by uniform random patterns, making the system more vulnerable to environmental changes.

The dispersal networks represent demographic connectivity between populations located in different habitats. These networks describe how offspring of species move between these habitats, forming connections that influence both the demography and the growth of populations \cite{mcmanus,mcmanus2}. Additionally, through the eco-evolutionary model, it was observed that random networks performed better in non-evolving populations, while regular networks favored populations with higher evolutionary potential \cite{mcmanus}. These networks, by reducing maladaptive gene flow, allowed local populations to adapt more efficiently. Results reinforce the importance of considering eco-evolutionary dynamics, network structures, and environmental gradients when assessing species' ability to migrate and persist under climate change.

\section{Data source and network}

\hspace{5mm}Data from the GeoSymbio \cite{franklin} and a complementary database \cite{williams} were used to construct the bipartite coral-symbiont network. Geosymbio database provided information about the organisms, such as \textit{Symbiodinium} type based on ITS2 sequence type, scientific name (genus and species) of coral and the location (i.e., ocean region) from which the \textit{Symbiodinium} specimen was collected. There is a total of 53 ocean regions in the GeoSymbio. The complementary database was used to obtain data on the thermal tolerances of the \textit{Symbiodinium} type and the coral host. Unfortunately, the database is not complete and some organism do not have a specified thermal tolerance. In such cases, the mean value of the thermal tolerance was used as reference. 

A bipartite network encoding the relationship between host corals and its endosymbiotic algae was generated for each ocean region. In particular, an edge represents the symbiotic relationship between a symbiont species and a host species in the ocean region where it were observed (Fig. \ref{fig:network}). Thus, there are no edges between organisms of distinct regions. Besides, each node represents a symbiont species or host species in a region. For instance, if a same species of symbiont or host occurs in $k$ regions, then the network will have $k$ vertices of this species.

\vspace{-5mm}

\definecolor{myblue}{RGB}{104,149,197}
\definecolor{myyellow}{RGB}{255,243,128}
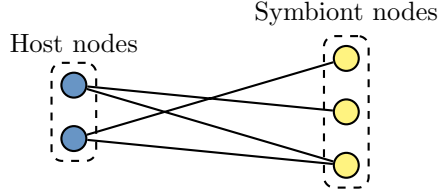
\begin{figure}[H]
    \centering
\begin{tikzpicture}[thick,amat/.style={matrix of nodes,nodes in empty cells,
  row sep=1em,draw,dashed,rounded corners,
  nodes={draw,solid,circle,execute at begin node={}}},
  fsnode/.style={fill=myblue},
  ssnode/.style={fill=myyellow}]

 \matrix[amat,nodes=fsnode,label=above:Host nodes] (mat1) {\\
 \\};

 \matrix[amat,right=3cm of mat1,nodes=ssnode,label=above:Symbiont nodes ] (mat2) {\\
 \\ 
 \\};

 \draw  (mat1-1-1) -- (mat2-2-1)
        (mat1-1-1) -- (mat2-3-1)
        (mat1-2-1) -- (mat2-3-1)
        (mat1-2-1) -- (mat2-1-1);
\end{tikzpicture}
\caption{Bipartite coral–symbiont network with host nodes (blue) and symbiont nodes (yellow).} \label{fig:network}
\end{figure}

The global coral–symbiont network has 867 symbiont nodes and 1178 host nodes, 2747 edges and 181 connected components. Note that the connected components are at least the number of ocean regions (i.e., 53), however the global network has many more connected components. Hence, there are multiple connected components within the same ocean region. 

Moreover, five connected components of the global network each corresponding to a different region were chosen to be analyzed separately, as shown in Table \ref{tab:components}. These regions represent the most threatened regions of coral bleaching in the oceans. Note that these networks have different number of nodes and edges, but relatively similar edge density.

\vspace{-5mm}

\setlength{\tabcolsep}{5pt} 
\renewcommand{\arraystretch}{1.2} 
\begin{table}[H]
\centering
\caption{Network nodes and edges in different connected components analyzed.}\label{tab:components}
 \begin{tabular}{||c c c c c||} 
 \hline
 Region & Symbiont nodes & Host nodes &  Edges & Density\\ [0.5ex] 
 \hline\hline
 Great Barrier Reef & 76 & 198 & 415 & 0.055\\ 
 Phuket & 36 & 152 & 442 & 0.162\\
 Western Indian & 43 & 131 & 337 & 0.120\\
 Western Caribbean  & 36 & 61 & 111 & 0.101\\ 
 Florida  & 26 & 32 & 75 & 0.180\\[1ex] 
 \hline
 \end{tabular}
\end{table}

\vspace{-5mm}

\subsection{Degree distribution}

\hspace{5mm} The degree of the global coral-symbiont network is analyzed, considering all 53 ocean regions. Table~\ref{tab:degree} shows that the average degree of both types of nodes is relatively similar but not the standard deviation which is larger for the symbiont nodes. Furthermore, since the minimum degree is 1, there are no isolated organisms (nodes) in the coral-symbiont network.

\vspace{-5mm}

\setlength{\tabcolsep}{5pt} 
\renewcommand{\arraystretch}{1.2} 
\begin{table}[H]
\centering
\caption{Degree in the global coral–symbiont network.}\label{tab:degree}
 \begin{tabular}{||c c c c c||} 
 \hline
 Type of & Standard & Minimum & Average & Maximum \\ 
 node & Deviation & degree & degree & degree\\ [0.5ex] 
 \hline\hline
 Symbiont & 8.306 & 1 & 3.168 & 102 \\ 
 Host & 2.346 & 1 & 2.332 & 51\\
 [1ex] 
 \hline
 \end{tabular}
\end{table}

\vspace{-5mm}

\begin{figure}[H]
    \centering
    \includegraphics[width=0.8\linewidth]{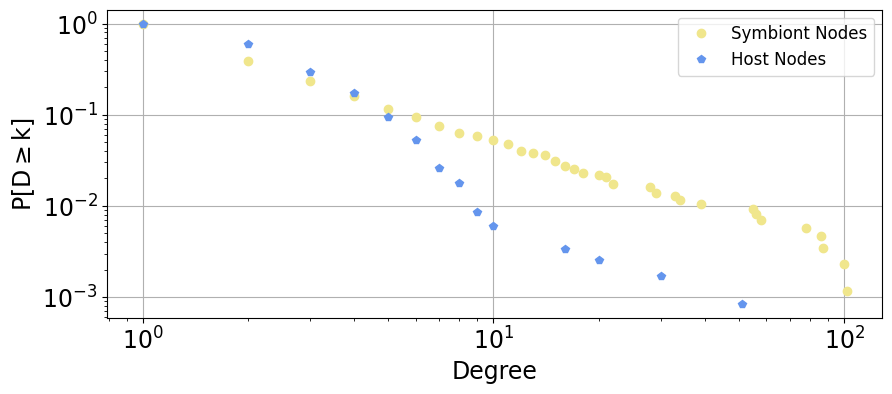}
    \caption{The complementary cumulative distribution function (CCDF) of  symbiont and host nodes.}
    \label{fig:ccdfdegree}
\end{figure}

Fig.~\ref{fig:ccdfdegree} shows the complementary cumulative distribution function (CCDF) for the degree of both symbiont and host nodes. Note that both distributions are heavy-tailed since a tiny number of symbionts are connected 100 or more hosts and a tiny number of hosts are connected to 50 or more symbionts. Further, note that the symbionts have a heavier tail (the distribution curve decreases more slowly) indicating that symbionts connect more, also because the number of host nodes is much larger.  

\vspace{-5mm}

Moreover, the difference between the tail values and the average degree values of the two types of nodes is very significant. Recall that the average degree of the symbionts and hosts are approximately 3 and 2, respectively. Thus, the majority of hosts and symbionts are specialists (have very few connections) while a tiny amount of both nodes are generalists (have large number of connections).

\section{Mathematical model for population growth}

\hspace{5mm}A novel population model using the network, thermal resistance, and symbiotic population was developed with the aim of studying the population dynamics of coral and algae under exposure to recurrent warming events. In what follows the model is described in detail and Table \ref{tab:symbols} presents variables and parameters of the model.

\vspace{-5mm}

\setlength{\tabcolsep}{5pt} 
\renewcommand{\arraystretch}{1.2} 
\begin{table}[H]
\centering
\caption{Definition for symbols of variables and parameters of the model.} \label{tab:symbols}
 \begin{tabular}{||c l||} 
 \hline
 Symbol & Definition (variables and parameters) \\ [0.5ex] 
 \hline\hline
 $S_i(t)$ & population of the $i$-th symbiont species at time $t$\\ 
 $H_i(t)$ & population of the $i$-th host species at time $t$\\ \hline 
 $N_{i}^{s}$  & neighborhood of the $i$-th symbiont species\\
 $N_{i}^{h}$  & neighborhood of the $i$-th host species \\ 
 $r_{i}^{s}$  & population growth rate of the $i$-th symbiont species\\
 $r_{i}^{h}$  & population growth rate of the $i$-th host species \\
 $m_{i}^{s}$  & population mortality rate of the $i$-th symbiont species\\
 $m_{i}^{h}$  & population mortality rate of the $i$-th host species \\
 $\tau_{i}^{s}$ & thermal tolerance of the $i$-th symbiont species\\
 $\tau_{i}^{h}$ & thermal tolerance of the $i$-th host species \\
 [1ex] 
 \hline
 \end{tabular}
\end{table}

In essence, the  model is a system of coupled ordinary differential equations to track the population of symbionts and hosts over time. This model considers the coral-symbiont network, where every node has associated with it a population. Note that this model uses a single variable per node instead of a variable for each symbiotic relationship (i.e, edges). Consequently, this model has significantly fewer variables (see Table \ref{tab:components}). However, network edges drive the population dynamics as growth of corals and algae are coupled and symbiotic. 

Let $S_i(t)$ and $H_i(t)$ denote the population of symbiont $i$ and host $i$ at time $t$, respectively. The evolution (derivative) of $S_i$ over time is given by: 
\begin{equation}
\frac{dS_i}{dt} = 
\frac{S_i}{|N_{i}^{s}|}r_{i}^{s}\left(\sum_{j \in |N_{i}^{s}|} \space\frac{H_j}{|N_{j}^{h}|}\right) - S_i m_{i}^{s}
\label{eq:S_i}
\end{equation}

Note that there is a growth term (positive) and a mortality term (negative) that are driven by a growth rate ($r_{i}^{s}$)  and mortality rate ($m_{i}^{s}$). Moreover, the growth term also depends on the network. This is the main contribution of the proposed model. In particular, the growth rate depends on the population of the corals that have a symbiotic relationship (edge) with this symbiont. 

In particular, the growth rate is multiplied by the sum across the neighboring hosts of the fraction of the host populations ($H_j$) divided by its neighbors ($N_{j}^{h}$). This fraction is assumed to interact with a fraction of this symbiont population, which is given by $S_i$ divided by its neighbors ($N_{i}^{s}$). Thus, for each neighboring host $j$, the growth rate is multiplied by $\displaystyle\frac{H_j}{N_{j}^{h}} \frac{S_i}{N_{i}^{s}}$. Note that the second term does not depend on $j$. 

Dividing the population of an organism by its degree assumes that each population interacts uniformly with the population of neighboring organism. This normalization ensures that the interaction of a symbiont or host population is distributed evenly among its connections. While hosts typically have fewer neighbors than symbionts, this asymmetry is inherent to the network structure and is represented in the model by this normalization. Moreover, this assumption significantly simplifies the model as it requires a single variable (population) for each node while also capturing network heterogeneity (different degrees).

The growth rate ($r_{i}^{s}$) is given by:
\begin{equation}
 r_{i}^{s} = \displaystyle \frac{r_{0}^{s}}{\sqrt{2\pi\left(\tau_{i}^{s}\right)^2}}.e^{\left(\frac{-(T(t) - z)^2}{\left(\tau_{i}^{s}\right)^2}\right)}
 \label{eq:growth_rate}
\end{equation}

While the mortality rate ($m_{i}^{s}$) is given by: 
\begin{equation}
m_{i}^{s} = \begin{cases}
\mu , &\text{ if $T(t) \leq z$}\\\vspace{2mm}\\
\displaystyle 1 -  e^{\left(\frac{-(T(t) - z)^2}{\left(\tau_{i}^{s}\right)^2}\right)},&\text{ if $T(t) > z$}
\end{cases}
\label{eq:mortality_rate}
\end{equation}

Note that both the growth and mortality rates have already been proposed in the literature \cite{mcmanus,walsworth} and depend on the current local sea temperature ($T(t)$) and thermal tolerance ($\tau_{i}^{s}$) of each organism. 

Considering the mortality rate, note that if the temperature is lower than or equal to the optimum temperature for growth (given in the model by parameter $z$), the mortality rate is equal to $\mu$ (see value in Table \ref{tab:parameters}). However, if the current local sea temperature is higher than the ideal growth temperature, the mortality rate is a function that depends on the thermal tolerance of the organisms.

The evolution (derivative) of $H_i$ over time is given by: 
\begin{equation}
\frac{dH_i}{dt} = 
\frac{H_i}{|N_{i}^{h}|} r_{i}^{h} \left(\sum\limits_{j \in N_{i}^{h}} \frac{S_j}{|N_{j}^{s}|} \right) - H_i m_{i}^{h}
\label{eq:H_i}
\end{equation}

Note that this equation is identical to Eq. \ref{eq:S_i} making the model symmetric. The growth rate and mortality rate for hosts are also given by equations Eq. \ref{eq:growth_rate} and Eq. \ref{eq:mortality_rate}, respectively (replacing superscript $s$ with $h$, as shown in Table~\ref{tab:symbols}). Thus, there is no inherent population growth advantage between symbionts and hosts. Of course, their growth depends on the parameters of the model such as network structure, thermal tolerance, water temperature and initial population.

\vspace{-5mm}

\setlength{\tabcolsep}{5pt} 
\renewcommand{\arraystretch}{1.2} 
\begin{table}[H]
\centering
\caption{Parameter definitions and values used in simulations.}
\label{tab:parameters}
 \begin{tabular}{||c c c||} 
 \hline
 Parameter & Value & Definition \\ [0.5ex] 
 \hline\hline
 $r_{0}^{s}$ & 1.0 & scaling factor for symbionts' growth rate \cite{mcmanus} \\ 
 $r_{0}^{h}$ & 1.0 & scaling factor for hosts' growth rate \cite{mcmanus} \\
 $z$ & 29.1°C & optimum growth temperature for symbionts and coral hosts\\
 $\mu$ & 0.3 & the base mortality \cite{walsworth} \\ [1ex] 
 \hline
 \end{tabular}
\end{table}
 
The growth and mortality rate of symbionts and hosts depend on the current local sea temperature. Thus, a model for the evolution of the sea temperature is needed. The temperature model used in this work was based on real ocean temperature data, collected over 26 months, in two regions of Western Australia: Coral Bay and Tantabiddi \cite{tempmodel} which is shown to be recurrent. In particular, the temperature model is given by: 
\begin{equation}
T(t) = \displaystyle 4\text{cos}\left(\frac{t}{75} + 30.6\right) + 26
\end{equation}

\begin{figure}[H]
    \centering
            \includegraphics[width=0.6\linewidth]{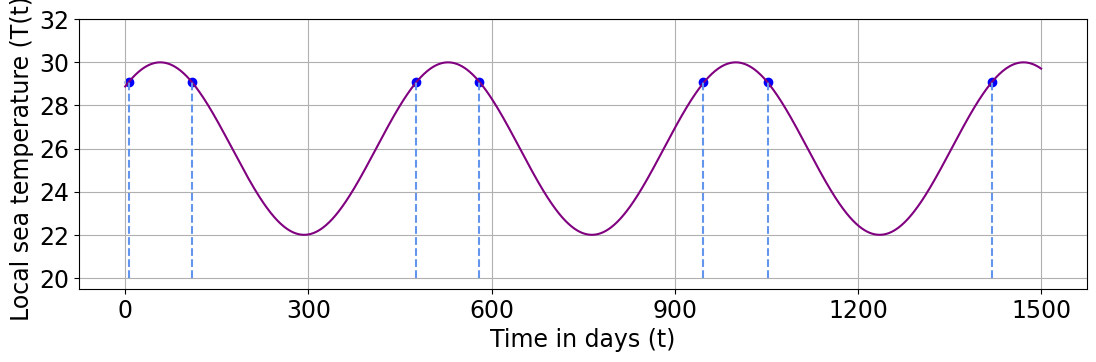}
    \caption{Local sea temperature function over time. The blue dashed lines represents the moment at which the model reached the optimal growth temperature ($z$).}\label{fig:temperature}
\end{figure}

The choice of parameters for the temperature model was arbitrary to emulate recurrence within a temperature range and timescale. Fig. \ref{fig:temperature} shows the evolution of the temperature over time indicating the optimal growth temperature value. 

\vspace{-5mm}

Finally, Eq. \ref{eq:S_i} and Eq. \ref{eq:H_i} will be solved numerically and independently for each region (see Table \ref{tab:components}) according to the above temperature model over a time horizon that simulates successive warming events over 1500 days, as shown in Fig. \ref{fig:temperature}.

\section{Quantitative analysis}

\hspace{5mm} Numerical solution of the population model provides insights into how populations of host corals and endosymbiotic algae behave when exposed to successive warming events when their growth is coupled by the network. Moreover, assuming that all host and symbiont species have some initial population, it is possible to characterize the role of the symbiotic interactions network structure in the growth dynamics of these populations and how the the network influences recovery after warming events. 

In particular, all symbiont species have an initial population of 1000, while all host species have an initial population of 100. Thus, there is no preferred species at time zero.

\subsection{Population growth}

\begin{figure}[H]
	\centering
	\begin{subfigure}[t]{0.49\linewidth}
		\includegraphics[width=\linewidth]{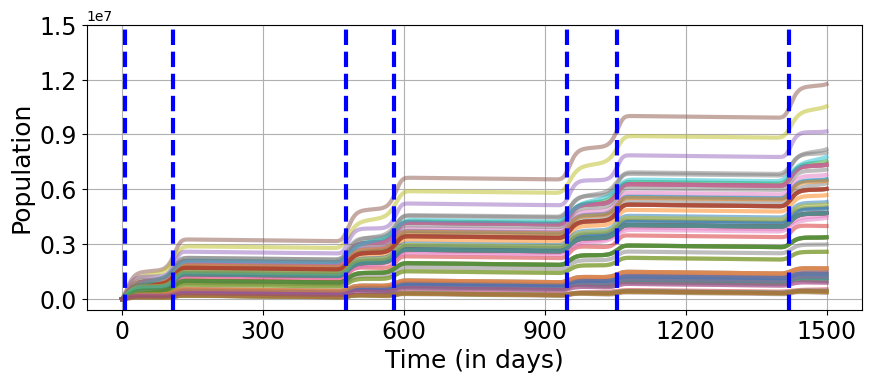}
		\caption{Great Barrier Reef - Symbionts' growth}
	\end{subfigure}
\hfill
	\begin{subfigure}[t]{0.49\linewidth}
		\includegraphics[width=\linewidth]{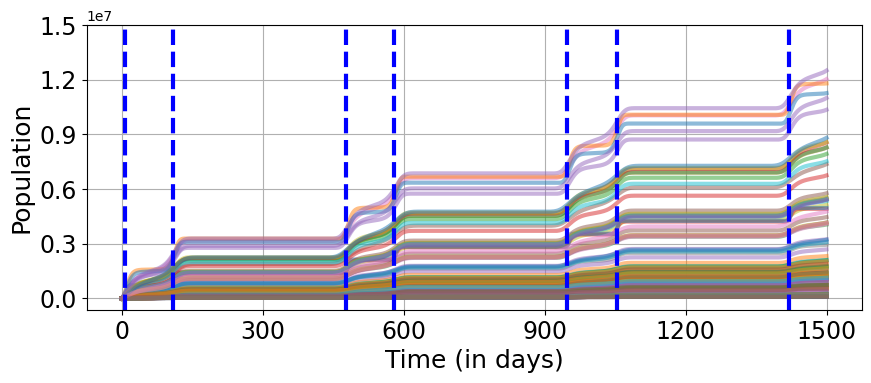}
		\caption{Great Barrier Reef - Hosts' growth}
	\end{subfigure}\\
	\caption{Population dynamics of symbiont species (a) and host species (b) at Great Barrier Reef over time. Blue dashed lines indicate the moment when the optimum temperature ($z$) for growth was reached.}
	\label{fig:gbrgrowth}
\end{figure}

\hspace{5mm} Fig. \ref{fig:gbrgrowth} shows the population growth for symbionts and hosts at the Great Barrier Reef region. Note that all species showed an overall increasing trend in the population.

Moreover, when water temperature is far from the optimal the population of most species decreases. This same trend was observed in all other regions. However, all species showed resilience as they continue to grow in population despite the thermal stress events.

\begin{figure}[H]
	\centering
	\begin{subfigure}[t]{0.49\linewidth}
		\includegraphics[width=\linewidth]{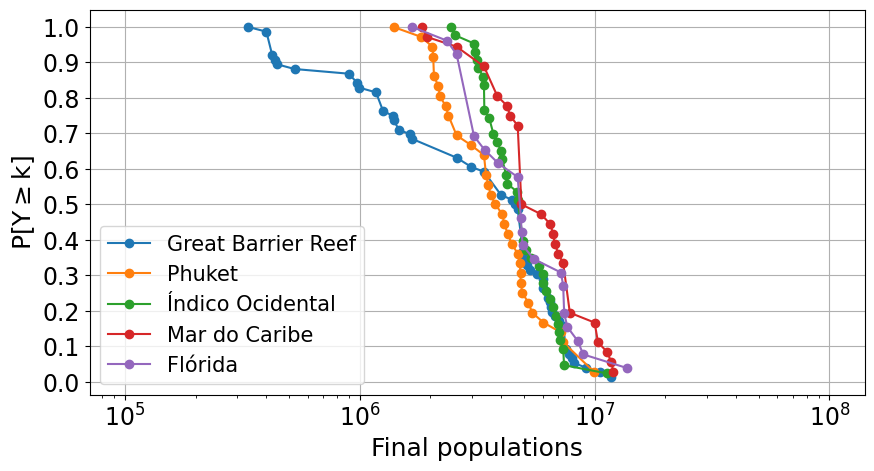}
		\caption{CCDF of symbionts' final populations}
	\end{subfigure}
\hfill
	\begin{subfigure}[t]{0.49\linewidth}
		\includegraphics[width=\linewidth]{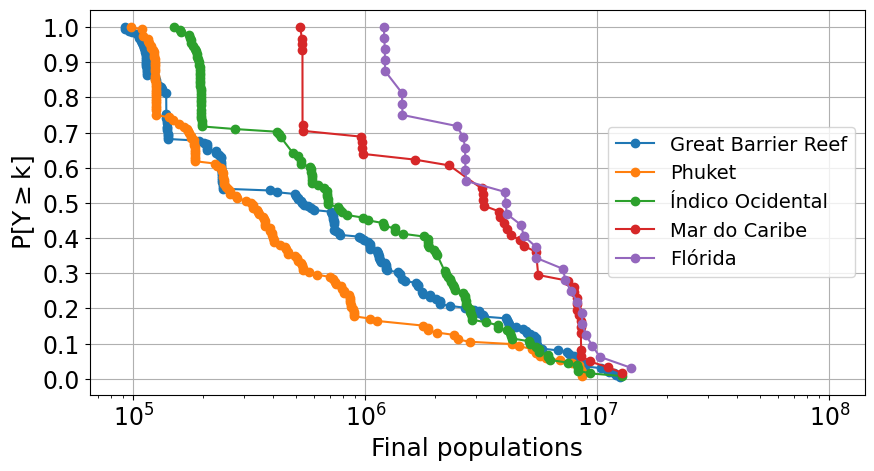}
		\caption{CCDF of hosts' final populations}
	\end{subfigure}\\
	\caption{CCDFs of symbionts' (left) and hosts' (right) final populations in their respective collection region.}
 \label{fig:ccdfpopfinalreal}
\end{figure}

\vspace{-5mm}

Nevertheless, even with the initial populations being the same for all species, differences in the evolution of populations occur due to the structure of the network and thermal resistances. Since the network structure is not uniform, as the node degrees are very different, population growth is also not uniform. Fig.~\ref{fig:ccdfpopfinalreal} shows the population distribution after 1500 days for both symbionts and hosts for all regions analyzed. Note that population distribution for hosts exhibits a heavy tail in all regions.This highlights the central role of network structure in population dynamics, as it determines how the species interactions shape resilience. Species with higher node degree (generalist species) adapt better to environmental changes, while species with lower node degree (specialist species) grow slower when exposed to thermal disturbances.

\begin{figure}[H]
	\centering
	\begin{subfigure}[t]{0.45\linewidth}
		\includegraphics[width=\linewidth]{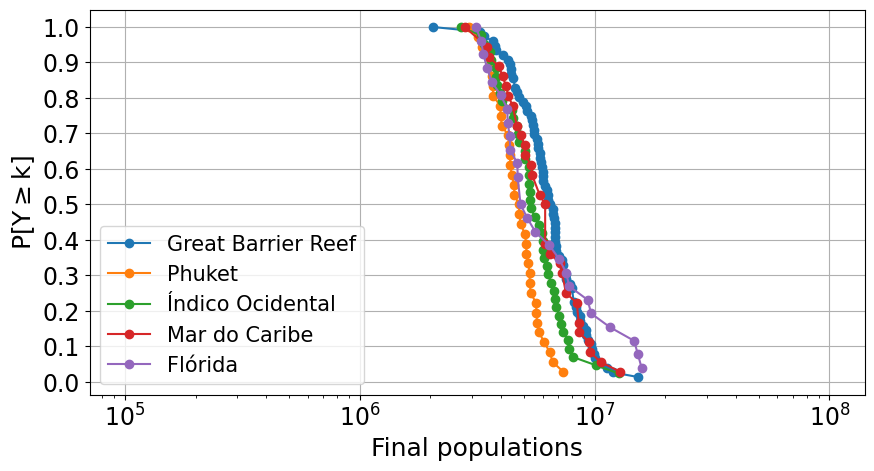}
		\caption{CCDF of symbionts' final populations}
	\end{subfigure}
\hfill
	\begin{subfigure}[t]{0.45\linewidth}
		\includegraphics[width=\linewidth]{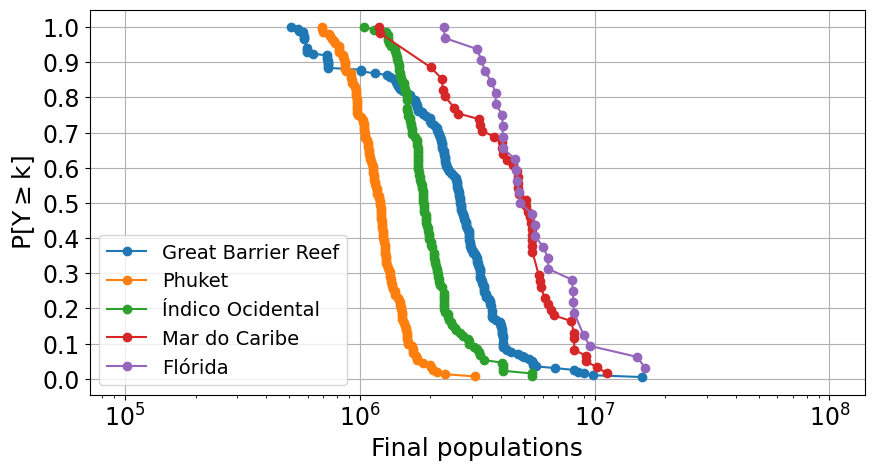}
		\caption{CCDF of hosts' final populations}
	\end{subfigure}\\
	\caption{CCDFs of symbionts' (left) and hosts' (right) final populations in their respective collection region in the random networks.}
 \label{fig:ccdfpopfinalrandom}
\end{figure}

\subsection{Influence of network structure}

\hspace{5mm}In order to study whether the structure of interactions (network) influences the resistance of species to thermal stresses, a random bipartite network was created for each region (as shown in Tab.~\ref{tab:components}) using a previously described methodology~\cite{williams}. In particular, each original edge was repositioned uniformly at random, destroying any biological symbiotic affinity. Note that the number of edges of each network was preserved. Moreover, the network randomization procedure adopted did not allow any isolated nodes, as all nodes in the randomized network have degree of at least 1. Fig. \ref{fig:ccdfpopfinalrandom} shows the population distribution for each region when growing on the random networks. Interestingly, the distributions have a much lighter tail in comparison to the original networks (see Fig.~\ref{fig:ccdfpopfinalreal}). 

The role of the network structure and the thermal tolerances of species on the population can be studied through correlation analysis. 

Table \ref{tab:correlation_thermal} shows the correlation between the thermal tolerances of species and their final populations. Note that this value is relatively low across all regions (close to zero). Therefore, final population are not even moderately correlated with the thermal tolerance.

\vspace{-6mm}

\setlength{\tabcolsep}{5pt} 
\renewcommand{\arraystretch}{1.2} 

\begin{table}[H]
\centering
\caption{Correlation between final populations and thermal tolerances.}\label{tab:correlation_thermal}
 \begin{tabular}{||c c c||} 
 \hline
 Region & Symbiont nodes & Host nodes\\ [0.5ex] 
 \hline\hline
 Great Barrier Reef & 0.134 & -0.085 \\ 
 Phuket & -0.008 & -0.022 \\
 Western Indian & 0.133 & -0.130 \\
 Western Caribbean  & 0.205 & 0.132 \\ 
 Florida  & 0.263 & -0.151 \\[1ex] 
 \hline
 \end{tabular}
\end{table}

\vspace{-1.2cm}

\setlength{\tabcolsep}{5pt} 
\renewcommand{\arraystretch}{1.2} 
\begin{table}[H]
\centering
\caption{Correlation between final populations and node degrees. For 
the random network, the sample average correlation and its standard deviation is reported using 50 independent instances of the random network.
}\label{tab:correlation_degree}
 \begin{tabular}{||c  c | c||} 
 \hline
 Region & Real network & Random network \\ [0.5ex] 
 \hline\hline
     & Symbionts \quad   Hosts & \hspace{-1mm}Symbionts \hspace{9mm}   Hosts\\
 Great Barrier Reef & \hspace{3mm} 0.384 \hspace{6mm} 0.552  & \hspace{2mm}0.213 $\pm$ 0.072 \hspace{1mm} 0.168 $\pm$ 0.073\\ 
 Phuket & \hspace{3mm} 0.193 \hspace{6mm} 0.186 & \hspace{2mm}0.232 $\pm$ 0.144 \hspace{1mm} 0.078 $\pm$ 0.087\\
 Western Indian & \hspace{3mm} 0.183 \hspace{6mm} 0.206 & \hspace{2mm} 0.075 $\pm$ 0.164 \hspace{1mm} 0.100 $\pm$ 0.092\\
 Western Caribbean  & \hspace{3mm} 0.469 \hspace{6mm}  0.579 & \hspace{2mm}0.154 $\pm$ 0.125 \hspace{1mm} 0.146 $\pm$ 0.134\\ 
 Florida  & \hspace{3mm} 0.383 \hspace{6mm} 0.404 & \hspace{2mm}0.078 $\pm$ 0.216 \hspace{1mm} 0.084 $\pm$ 0.202\\[1ex] 
 \hline
 \end{tabular}
\end{table}

In contrast, Table \ref{tab:correlation_degree} shows the correlation between the final population and node degree. Note that this correlation is relatively larger than with thermal tolerance for all ocean regions. Moreover, for three regions the correlation is above 0.3 (considered a moderate value) for both symbionts and hosts. In biological terms, this indicates that symbionts and corals that are generalists (have higher degree) are able to growth faster than those that are specialists (have lower degree) in the presence of water temperature variation.

Table \ref{tab:correlation_degree} also shows the correlation between degree and final population in the random networks. Note that the average correlation for the random networks is considerably smaller than the original networks that have moderate correlation (above 0.3). As expected, randomly repositioning the edges removes the heavy tail property and makes the degree distribution more centered. These results reinforce the importance of network structure for the survival of these organisms.

\begingroup
\setlength{\tabcolsep}{5pt} 
\renewcommand{\arraystretch}{1.2} 
\begin{table}[H]
\centering
\caption{Correlation between final populations and sum of degrees of neighbors.}\label{tab:neighbor_degree}
 \begin{tabular}{||c  c  c||} 
 \hline
 Region & Symbiont nodes & Host nodes \\ [0.5ex] 
 \hline\hline
 Great Barrier Reef & 0.181 & -0.235\\ 
 Phuket &  0.132 & -0.365\\
 Western Indian & 0.127 & -0.499 \\
 Western Caribbean  & 0.308 & -0.229 \\ 
 Florida  & 0.105 & -0.238 \\[1ex]
 \hline
 \end{tabular}
\end{table}
\endgroup

Finally, Table \ref{tab:neighbor_degree} shows the correlation between the sum of degrees of neighbors and the final populations of each species. Differently from degree correlation, negative correlations values for the hosts stand out. Note that, in all analyzed networks, host nodes are more numerous (Table \ref{tab:components}). Therefore, symbiont nodes have many more neighbors than host nodes (Fig. \ref{fig:ccdfdegree}). Thus, in Eq. 4, the growth of the hosts, determined by $r_{i}^h$ and network structure, is harmed by the large number of neighbors of the symbionts, since the sum of fractions are smaller due to their larger denominator ($|N_{j}^s|$). Hence, the lower the degree of the hosts' neighbors, the larger the sum fractions (Eq. \ref{eq:H_i}) in the contribution to the growth rate of the hosts. This relationship is an indicative of the negative correlation between these two variables.

On the other hand, this correlation for symbionts is always positive, although weak, since their neighbors tend to have smaller degrees which will increases their growth rate, determined by $r_{i}^s$ and network structure. Thus, correlation of neighbors degrees is not symmetric between hosts and symbionts, differently from degree (where three regions had moderate correlation for both hosts and symbionts).

\section{Conclusion}

\hspace{5mm}This paper investigated the population dynamics within the global coral–sym-
biont network under temperature variations, with a focus on the impact of thermal stress on coral bleaching. Using a bipartite network model, the relationships between coral hosts and their symbiotic algae has been characterized, identifying how network structure influences population growth and resilience to recurrent warming events. Besides the numerical analysis, a main contribution of this work is a simple and parameterized mathematical model capturing the network structure. 

Our results demonstrated that the network structure plays a crucial role in determining the capacity of coral and symbiont species to recover from warming events, with generalist species exhibiting stronger recovery patterns. 

Furthermore, correlations between final populations and node degrees emphasized the importance of network connectivity in population growth. These findings enhance the understanding of the ecological factors that contribute to coral reef resilience and underscore the need to consider network structure when evaluating species’ adaptability to climate change.

\bibliographystyle{plain}
\bibliography{ms}

\begin{thebibliography}{1}

\bibitem{donner}
Simon~D Donner, William~J Skirving, Christopher~M Little, Michael Oppenheimer, and OVE Hoegh-Guldberg.
\newblock Global assessment of coral bleaching and required rates of adaptation under climate change.
\newblock {\em Global Change Biology}, 11(12):2251--2265, 2005.

\bibitem{franklin}
Erik~C Franklin, Michael Stat, Xavier Pochon, Hollie~M Putnam, and Ruth~D Gates.
\newblock Geosymbio: a hybrid, cloud-based web application of global geospatial bioinformatics and ecoinformatics for symbiodinium--host symbioses.
\newblock {\em Molecular Ecology Resources}, 12(2):369--373, 2012.

\bibitem{tempmodel}
Christopher~J Fulton, Martial Depczynski, Thomas~H Holmes, Mae~M Noble, Ben Radford, Thomas Wernberg, and Shaun~K Wilson.
\newblock Sea temperature shapes seasonal fluctuations in seaweed biomass within the ningaloo coral reef ecosystem.
\newblock {\em Limnology and Oceanography}, 59(1):156--166, 2014.

\bibitem{hughes}
Terry~P Hughes, James~T Kerry, Andrew~H Baird, Sean~R Connolly, Andreas Dietzel, C~Mark Eakin, Scott~F Heron, Andrew~S Hoey, Mia~O Hoogenboom, Gang Liu, et~al.
\newblock Global warming transforms coral reef assemblages.
\newblock {\em Nature}, 556(7702):492--496, 2018.

\bibitem{mcmanus2}
Lisa~C McManus, Daniel~L Forrest, Edward~W Tekwa, Daniel~E Schindler, Madhavi~A Colton, Michael~M Webster, Timothy~E Essington, Stephen~R Palumbi, Peter~J Mumby, and Malin~L Pinsky.
\newblock Evolution and connectivity influence the persistence and recovery of coral reefs under climate change in the caribbean, southwest pacific, and coral triangle.
\newblock {\em Global Change Biology}, 27(18):4307--4321, 2021.

\bibitem{mcmanus}
Lisa~C McManus, Edward~W Tekwa, Daniel~E Schindler, Timothy~E Walsworth, Madhavi~A Colton, Michael~M Webster, Timothy~E Essington, Daniel~L Forrest, Stephen~R Palumbi, Peter~J Mumby, et~al.
\newblock Evolution reverses the effect of network structure on metapopulation persistence.
\newblock {\em Ecology}, 102(7):e03381, 2021.

\bibitem{walsworth}
Timothy~E Walsworth, Daniel~E Schindler, Madhavi~A Colton, Michael~S Webster, Stephen~R Palumbi, Peter~J Mumby, Timothy~E Essington, and Malin~L Pinsky.
\newblock Management for network diversity speeds evolutionary adaptation to climate change.
\newblock {\em Nature Climate Change}, 9(8):632--636, 2019.

\bibitem{williams}
Sara~D Williams and Mark~R Patterson.
\newblock Resistance and robustness of the global coral--symbiont network.
\newblock {\em Ecology}, 101(5), 2020.

\bibitem{will_tese}
Sara~Dell Williams.
\newblock {\em Corals are more than the sum of their colonies: a network science perspective on the role of coral complexity and its consequences for coral reef health}.
\newblock PhD thesis, Northeastern University, 2020.

\end{thebibliography}

\end{document}